\documentclass[sigconf,screen]{acmart}

\usepackage{xspace}
\usepackage{multirow}
\usepackage{adjustbox}
\usepackage{xcolor}
\usepackage{listings}
\usepackage{balance}
\usepackage{tcolorbox}
\usepackage[inline]{enumitem}

\newcommand{\approach}{\mbox{\textsc{E-Test}}\xspace}

\newcommand{\tested}{\emph{already-tested}\xspace}
\newcommand{\untested}{\emph{not-yet-tested}\xspace}
\newcommand{\errorprone}{\emph{error-prone}\xspace}

\newcommand{\allScenarios}{\textsc{1,975}\xspace}

\newcommand{\cmpFastPlus}{\textsc{+20\%}\xspace}
\newcommand{\cmpFieldReady}{\textsc{+24\%}\xspace}
\newcommand{\tcgFailure}{\textsc{83.5\%}\xspace}

\newcommand{\preprocessor}{\textsc{PreProcessor}\xspace}
\newcommand{\analyzer}{\textsc{Analyzer}\xspace}
\newcommand{\postprocessor}{\textsc{PostProcessor}\xspace}

\definecolor{promptbgcolor}{gray}{0.9}
\lstdefinelanguage{prompt_template}{
    keywords={
        MUT,
        TEST,
        SUITE,
        SCENARIO,
        TASK,
        QUERIES
    },
    morecomment=[l]{//},
    basicstyle=\sffamily\tiny, % Set all text to tiny size
    keywordstyle=\color{teal}\textbf, % Keywords in blue
    showstringspaces=false, % Do not add space after string characters
    breaklines=true,         % Break lines automatically
    linewidth=\linewidth,           % Set line width for breaking
    columns=fullflexible
}

\AtBeginDocument{%
  }

\setcopyright{acmlicensed}
\setcopyright{cc}
\setcctype{by}
\begin{document}

%%
%% The "title" command has an optional parameter,
%% allowing the author to define a "short title" to be used in page headers.
\title{Ever-Improving Test Suite by Leveraging Large Language Models}

%%
%% The "author" command and its associated commands are used to define
%% the authors and their affiliations.
%% Of note is the shared affiliation of the first two authors, and the
%% "authornote" and "authornotemark" commands
%% used to denote shared contribution to the research.

\author{Ketai Qiu}
\orcid{0009-0002-9750-2762}
\affiliation{
  \institution{Faculty of Informatics, Universit\`a della Svizzera Italiana (USI)}
  \city{Lugano}
  \country{Switzerland}
}
\email{ketai.qiu@usi.ch}

%%
%% By default, the full list of authors will be used in the page
%% headers. Often, this list is too long, and will overlap
%% other information printed in the page headers. This command allows
%% the author to define a more concise list
%% of authors' names for this purpose.
\renewcommand{\shortauthors}{Qiu}

%%
%% The abstract is a short summary of the work to be presented in the
%% article.
\begin{abstract}
  Augmenting test suites with test cases that reflect the actual usage of the software system is extremely important to sustain the quality of long lasting software systems. 
 In this paper, we propose \approach, an approach that incrementally augments a test suite with test cases that exercise behaviors that emerge in production and that are not been tested yet.
 \approach leverages Large Language Models to identify \tested, \untested, and \errorprone unit execution scenarios, and augment the test suite accordingly. 
 Our experimental evaluation shows that \approach outperforms the main state-of-the-art approaches to identify inadequately tested behaviors and optimize test suites.
\end{abstract}

%%
%% The code below is generated by the tool at http://dl.acm.org/ccs.cfm.
%% Please copy and paste the code instead of the example below.
%%


%%
%% Keywords. The author(s) should pick words that accurately describe
%% the work being presented. Separate the keywords with commas.
%% A "teaser" image appears between the author and affiliation
%% information and the body of the document, and typically spans the
%% page.
% \begin{teaserfigure}
%   \includegraphics[width=\textwidth]{sampleteaser}
%   \caption{Seattle Mariners at Spring Training, 2010.}
%   \Description{Enjoying the baseball game from the third-base
%   seats. Ichiro Suzuki preparing to bat.}
%   \label{fig:teaser}
% \end{teaserfigure}

% \received{20 February 2007}
% \received[revised]{12 March 2009}
% \received[accepted]{5 June 2009}

%%
%% This command processes the author and affiliation and title
%% information and builds the first part of the formatted document.
\maketitle

\section{Introduction}

% Test suites are essential measures to guarantee the overall quality of software applications. Complex software systems without sufficient testing can cause disastrous consequences in production. For example, a single software update from CrowdStrike triggered global service outages from airlines to healthcare~\cite{Oliver:Forbes:2024}, resulting in a staggering \$5.4 billion market loss. This accident
% underscores the urgent need for robust testing strategies that can mitigate such risks.
% However, in reality, it is impossible to create a perfect test suite that can reveal all possible bugs in the application due to millions of combinations of testing scenarios~\cite{Gazzola:FieldFailures:ISSRE:2017}.
Test suites that sample the whole behavior of software systems in production can prevent sometime catastrophic failures, like the global service outages~\cite{Oliver:Forbes:2024} that was due to a single software update from CrowdStrike and that resulted in a staggering \$5.4 billion loss. 
Unfortunately, it is impossible to generate a perfect test suite that exercises all  scenarios and  prevents all failures in production~\cite{Gazzola:FieldFailures:ISSRE:2017, Denaro:PREFACE:FSE:2024}.

% While existing approaches focus on regression and field testing, there is a significant gap in efficiently maintaining test suites as the number of potential scenarios grows exponentially. Regression testing aims to continuously adapt the test suite as the software application evolves~\cite{Rothermel:RegressionTesting:TOSEM:2004, Shi:RegressionTestSelection:FSE:2015}. On the other hand, field testing approaches aims to reveal potential failures via executing emerging scenarios in production environment~\cite{Bertolino:ASurveyOfFieldBasedTestingTechniques:ACMCS:2021}.
Regression testing approaches maintain the test suites across incremental versions of the software systems by selecting and prioritizing the subsets of test cases to be re-executed on the new versions, and by relying on the former versions as test oracles~\cite{Rothermel:RegressionTesting:TOSEM:2004, Shi:RegressionTestSelection:FSE:2015}.  
%Field testing approaches enrich the test suites with new scenarios, by inferring test cases from production
%Regression testing approaches take advantage from the similarities across consecutive versions of software systems, and do not cope with the evolution of software's behaviors in production.
Field testing approaches leverage execution traces to test the actual behavior of software systems and enrich the test suites with scenarios from production~\cite{Bertolino:ASurveyOfFieldBasedTestingTechniques:ACMCS:2021, Qiu:ATEST:ICSE:2024}.

% The execution scenarios in production can be ideal sources to derive test cases since they are likely to explore not-already-tested scenarios. However, simply adding all scenarios into the test suite can quickly make its size explode, resulting in an unusable test suite. Instead, an effective approach involves continuously maintaining the test suite as the software evolves and progressively augmenting it with valuable testing scenarios. Specifically, we leverage Large Language Models (LLMs) to help quickly classify scenarios without execution based on the three elements of each scenario: the input data, the method under test and the corresponding test suite.
Ultimately, the set of execution traces observed in production is the best information about the actual use of a software system, since it includes all relevant scenarios, both the ones already tested and the ones that the test suites have missed so far.
However, simply adding all monitored scenarios to the test suite is clearly infeasible, due to the huge size of executions observed in production.

% In this paper, we propose \approach, an approach to \textit{continuously augment test suites by identifying not-yet-tested execution scenarios}. \approach classifies unit execution scenarios as \tested, \untested or \errorprone, and generates test cases with not-yet-tested scenarios to improve test suite quality. In particular, \approach leverages a fine-tuned LLM to classify unit execution scenarios through five queries and generate test cases via limited rounds of chat interactions. Our experimental evaluation demonstrates that \approach effectively classifies executions scenarios with an average F1-score of 54\%.
In this paper, we propose \approach, an effective approach to \textit{continuously augment test suites without dramatically inflating the size of the suite, by identifying the not-yet-tested execution scenarios, that is, the subset of scenarios that are observed in production and that require additional testing}. \approach monitors the execution of the software system at the unit level, classifies the scenarios monitored in production as \tested, \untested or \errorprone, 
and generates test cases from the not-yet-tested scenarios to improve the quality of the test suite.
\approach leverages a fine-tuned LLM to classify unit execution scenarios through five queries and generate test cases via limited rounds of chat interactions.

\section{Related Work}

% Our work is closely related with test suite augmentation, which focuses on enhancing existing test suites with valuable input data in regard to identify potential failures~\cite{Danglot:LiteratureTestAmplification:JSS:2019, Brandt:TestAmplification:ESE:2022, Alshahwan:AutomatedUnitTestImprovement:FSE:2024, Brandt:AmplifiedTests:TSE:2024}.
% \citet{Santelices:TestSuiteAugmentation:ASE:2008} proposed \textsc{MaTRIX} to identify inadequately tested behaviors in the test suite after the software update using symbolic execution, which is resource-demanding and is limited by the updated scope.
% \citet{Cruciani:ScalableTestSuiteReduction:ICSE:2019} implemented \emph{FAST++} approach to reduce the test suite size based on K-means clustering of test case vectors. \emph{FAST++} works only for granular class-wise test suites and ignores intrinsic code structures and relations.
% \citet{Shimmi:AutomatedTestSuiteMaintenance:ICST:2022} studied common patterns of co-evolution between source code and test suites to enable remediation of the test suite, which strongly relies on manually summarized patterns.
% \approach targets on improving an existing test suite with classified scenarios that examines the software consistently with its field usage and takes the complete codebase into account.
Test suite augmentation is an emerging research area in last few years, as a way to enhance test suites with new test cases that can identify potential failures~\cite{Danglot:LiteratureTestAmplification:JSS:2019, Brandt:TestAmplification:ESE:2022, Alshahwan:AutomatedUnitTestImprovement:FSE:2024, Brandt:AmplifiedTests:TSE:2024}.
% \citet{Santelices:TestSuiteAugmentation:ASE:2008} proposed \textsc{MaTRIX}, an approach that uses symbolic execution to identify behaviors that the test suite does not adequately exercise, after the software update.  The symbolic execution core of \textsc{MaTRIX} is resource-demanding, and \textsc{MaTRIX} is defined within the copse of a software update only.
\citet{Cruciani:ScalableTestSuiteReduction:ICSE:2019} defined \emph{FAST++}, an approach that uses K-means clustering of test case vectors to reduce the size of the test suite. \emph{FAST++} works only for granular class-wise test suites and does not consider intrinsic code structures and relations.
\citet{Shimmi:AutomatedTestSuiteMaintenance:ICST:2022} studied common patterns of co-evolution between source code and test suites, to enable the remediation of the test suite. Their approach strongly relies on manually summarized patterns.
\approach improves the test suite with new scenarios that emerge in production,  by excerpting the not-yet-tested scenarios from the execution traces using LLMs. \approach overcomes the limitations of state-of-the-art testing approaches by integrating a comprehensive analysis of source code, test cases, and execution traces thanks to strong code understanding capabilities learned from extensive training data.

% Field testing takes advantage of information extracted from the field to examine the software with unseen scenarios.
% \citet{Gazzola:FieldReadyTesting:ICST:2022} introduced field-ready testing, which executes emerging scenarios directly in production environment via instantiating parametric test case templates with data arising in production.
% \approach focuses on selecting scenarios from the field instead of executing the instantiated tests, which differs from field-ready testing.
Field testing leverages information extracted from the field using both static and dynamic analysis, to test the software systems with not-yet-tested  scenarios~\cite{Bertolino:ASurveyOfFieldBasedTestingTechniques:ACMCS:2021}.
\citet{Gazzola:FieldReadyTesting:ICST:2022} introduced \emph{field-ready testing}, which executes emerging scenarios directly in production environment via instantiating parametric test case templates with data arising in production.
\approach selects the scenarios that are observed in production and requires additional testing without actual execution, while \emph{field-ready testing} focuses on executing scenarios in production, that interferes with the running software.
\section{Approach}

\begin{figure}[t]
	\centering
	\includegraphics[width=\linewidth]{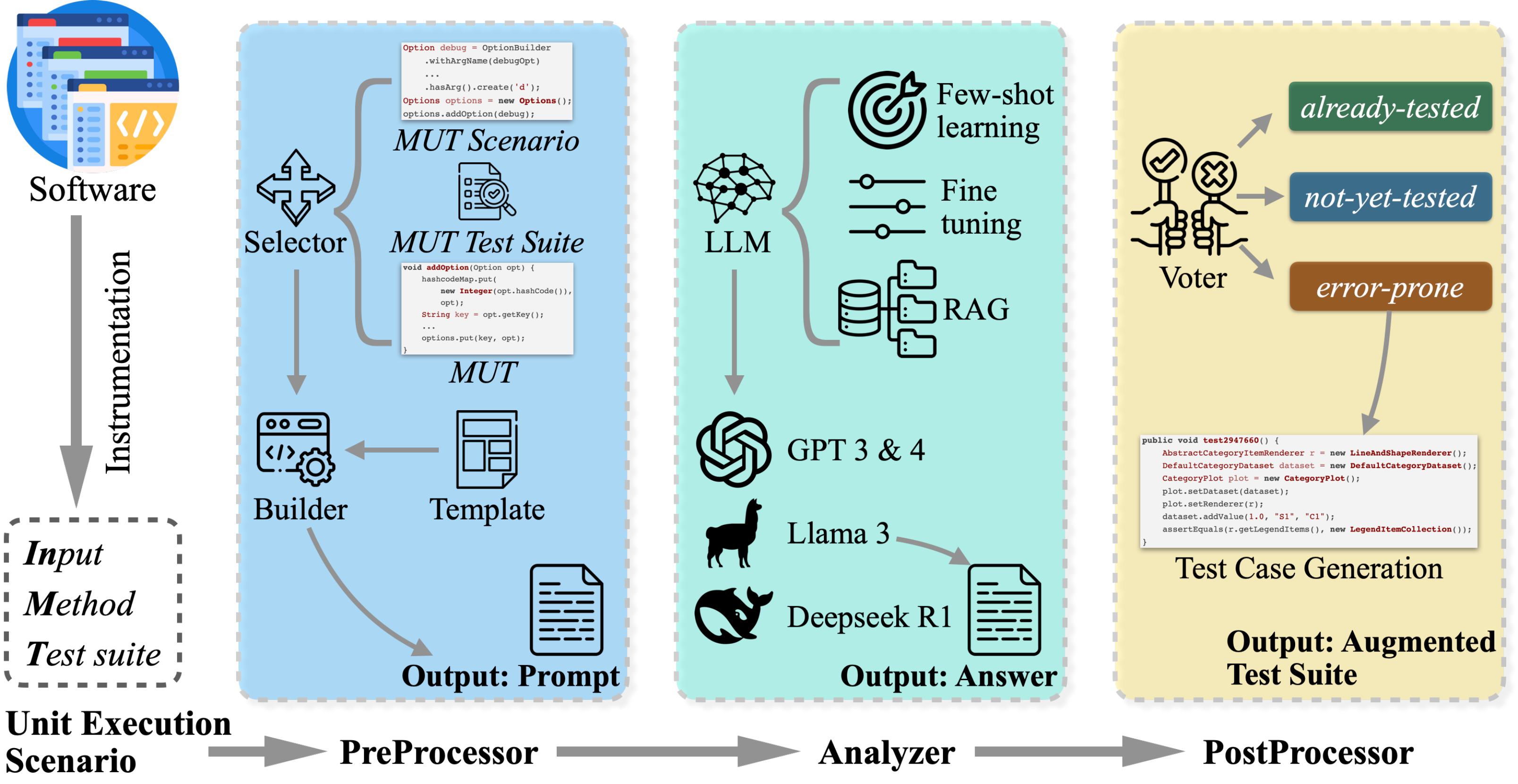}
	\caption{Overview of \approach.}
	\label{fig:overview}
\end{figure}

\approach monitors the input data $In$ of the focal methods $M$, by executing the instrumented focal methods in production, and classifies $In$ as \tested, \untested or \errorprone, with respect to a test suite $T$, in three steps: \preprocessor, \analyzer and \postprocessor as shown in Figure~\ref{fig:overview}. 
Specifically, we built a dataset from Defects4J and four Java projects on GitHub~\cite{GitHub} by instrumenting bug-revealing test cases. We provide the dataset and evaluation results as a replication package~\footnote{\url{https://github.com/ketaiq/etest-replication-package-fse25src}}.

The \preprocessor generates the prompt for the \analyzer by instantiating the template with five sections: \emph{MUT}, \emph{MUT Test Suite}, \emph{MUT Scenario}, \emph{Task} and \emph{Queries} as shown in Figure~\ref{fig:prompt_template}.
The five questions are formulated with the most suited terms about software bugs from Stack Overflow~\cite{StackOverflow}.

The \analyzer queries an LLM to answer the prompt with advanced configurations, including few-shot learning~\cite{Wang:SurveyFewshotLearning:ACMCS:2020}, Retrieval Augmented Generation (RAG)~\cite{Lewis:RetrievalAugmentedGeneration:NIPS:2020}, and fine-tuning~\cite{Weyssow:FineTuningCodeGenerationLargeLanguageModels:TOSEM:2025}.
We leveraged LlamaIndex~\cite{LlamaIndex} to build indices for all source code and tests in the examined Java project and fine-tuned GPT 3.5 Turbo using OpenAI APIs and 5\% of dataset.
We experimented the \analyzer with various LLMs using Ollama and OpenAI APIs, including Llama3 (8B and 70B), Deepseek-r1 70B, GPT (3.5 Turbo, 4 Turbo, 4o and fine-tuned 3.5 Turbo). 

% \postprocessor classifies the unit execution scenario as \tested, \untested or \errorprone by comparing the answer with the correct one for each query. The correct encodings of 3-class scenarios are: \texttt{10010} for \tested, \texttt{01010} for \untested and \texttt{01101} for \errorprone, where 1 represents \textsc{Yes} and 0 represents \textsc{No}. The classification is based on the vote of matched answers. For example, an answer encoding \texttt{10011} is classified as \tested as it has the highest number of matches. In case of a tie, a predefined order is applied: \errorprone, then \untested and finally \tested. Furthermore, \postprocessor can generate test cases for \untested or \errorprone scenarios via continuing the LLM chat interaction.
The \postprocessor classifies the input $In$ as \tested, \untested or \errorprone, by combining the answers of the \analyzer: \texttt{10010} = \tested, \texttt{01010} = \untested, \texttt{01101} = \errorprone,  where 1 and 0 corresponds to a \textsc{Yes} and \textsc{No} answer to the query, respectively. 
The classification depends on the best partial matching of the five answers. For example, the \postprocessor classifies an input $In$ with \texttt{10011} answers as \tested, by counting the matchings. In case of a tie, \errorprone dominates \untested that dominates \tested. 

\begin{figure}[t]
	\centering
    \begin{tcolorbox}[colback=promptbgcolor, boxrule=0pt, arc=5pt, left=1pt, right=1pt, top=0pt, bottom=0pt]
        \begin{lstlisting}[language=prompt_template]
MUT: // source code of the method under test 
MUT TEST SUITE: // a set of test cases for MUT
MUT SCENARIO: // monitored input data to MUT
TASK: Answer all questions in QUERIES. For each question, you should answer only YES or NO. Return results in a JSON dictionary.
QUERIES:
    Q1. Is MUT SCENARIO a similar scenario compared with MUT TEST SUITE?
    Q2. Does MUT SCENARIO cover more lines or branches than MUT TEST SUITE?
    Q3. Will MUT work differently when executed under MUT SCENARIO?
    Q4. Does MUT still produce correct results when executing under MUT SCENARIO?
    Q5. Will MUT SCENARIO reveal any bug in MUT?
        \end{lstlisting}
    \end{tcolorbox}
    \caption{Prompt template.}
    \label{fig:prompt_template}
\end{figure}
\section{Results}

We evaluated the effectiveness of \approach on a dataset of \allScenarios unit execution scenarios (537 \tested, 719 \untested, and 719 \errorprone scenarios) extracted from 17 Defects4J~\cite{Just:Defects4J:ISSTA:2014} projects and four popular open-source Java projects (Spring Boot~\cite{SpringBoot}, Apache ShardingSphere~\cite{ApacheShardingSphere}, Apache Dolphinscheduler~\cite{ApacheDolphinscheduler} and Micrometer~\cite{Micrometer}).
We split the Defects4J dataset into fine-tuning and validation sets with a 5:95 ratio. We fine-tuned GPT-3.5 Turbo with batch size 1, learning rate multiplier 2.0 and 3 epochs using the balanced fine-tuning dataset. We validated \approach on both the validation set from Defects4J and 150 scenarios from four GitHub projects.
% We evaluated the predictions of LLMs in terms of precision, recall and F1-score of each class as shown in Table~\ref{tab:effectiveness_metrics}.
% We refer to the fine-tuned GPT-3.5 Turbo as \approach.
% We computed the average F1-score across 3 classes to measure the overall effectiveness of each LLM.
% The average F1-scores of prompting LLMs range between 24\% and 32\%, which is worse than prompting with RAG (-7\% in terms of Llama3 8B). Few-shot prompting using GPT-4 Turbo did not improve significantly over zero shot (only +2\% using 3 shots).
% \approach demonstrates an average F1-score of \avgFscore for 3-class classification of scenarios, which significantly outperforms two state-of-the-art approaches: \approach achieves a \cmpFastPlus improvement over \textit{FAST++}~\cite{Cruciani:FASTTestSuiteReduction:ICSE:2019} and a \cmpFieldReady improvement over \textit{field-ready testing}~\cite{Gazzola:FieldReadyTesting:ICST:2022}. 
We measured the effectiveness of \approach as the amount of inputs that \approach correctly classifies as \tested, \untested or \errorprone, with respect to the available test suites. We compared \approach (instantiated on fine-tuned GPT-3.5 Turbo) with different LLMs both with and without RAG, and with the two closest approaches \emph{FAST++}~\cite{Cruciani:ScalableTestSuiteReduction:ICSE:2019} and \emph{field-ready testing}~\cite{Gazzola:FieldReadyTesting:ICST:2022}. 
Table~\ref{tab:effectiveness_metrics} reports the precision, recall and F1-score of each class for  \approach with and without RAG (last two rows), the LLMs (top rows in the table), and \emph{FAST++} and \emph{field-ready testing} (rows \citet{Cruciani:ScalableTestSuiteReduction:ICSE:2019} and \citet{Gazzola:FieldReadyTesting:ICST:2022}). 
\approach performs significantly better than the other LLMs, and largely outperforms both \emph{FAST++} (\cmpFastPlus) and \emph{field-ready testing} (\cmpFieldReady).
Furthermore, a case study on test case generation reveals that the identified \errorprone scenarios can detect \tcgFailure of failures. The experimental results demonstrate \approach's significant potential to enhance both test suite quality and software reliability.

\begin{table}[t]
	
	\caption{Precision (P), Recall (R) and F1-score (F1) of 3-class classification of scenarios using the experimented LLMs (\%)}
	\adjustbox{max width=\linewidth}{
        \centering
	\begin{tabular}{l|rrr|rrr|rrr|r}
		\toprule
		\multirow{2}{*}{\textbf{LLM}} & \multicolumn{3}{c|}{\textbf{\tested}} & \multicolumn{3}{c|}{\textbf{\untested}} & \multicolumn{3}{c|}{\textbf{\errorprone}} & \multirow{2}{*}{\textbf{Avg. F1}} \\
		& \textbf{P} & \textbf{R} & \textbf{F1} & \textbf{P} & \textbf{R} & \textbf{F1} & \textbf{P} & \textbf{R} & \textbf{F1} & \\ \midrule
		Llama3 8B & 26 & 19 & 22 & 36 & 51 & 42 & 36 & 27 & 31 &  32 \\
		Llama3 70B & 18 & 29 & 22 & 34 & 28 & 31 & 28 & 19 & 23 &  25 \\
		Deepseek-r1 70B & 18 & 29 & 22 & 29 & 24 & 26 & 52 & 45 & 48 & 32 \\
		GPT-3.5 Turbo & 26 & 10 & 15 & 35 & \textbf{64} & 46 & 36 & 23 & 28 & 30 \\
		GPT-4 Turbo & 17 & 20 & 18 & 34 & 56 & 42 & 31 & 7 & 11 & 24 \\
		GPT-4o & 19 & 33 & 24 & 32 & 38 & 16 & 37 & 10 & 35 & 25 \\
		\midrule
		GPT-4 Turbo (3-shot) & 19 & 24 & 21 & 35 & 52 & 42 & 32 & 9 & 14 & 26 \\
		GPT-4 Turbo (6-shot) & 15 & 19 & 17 & 35 & 55 & 43 & 32 & 8 & 13 & 24 \\
		GPT-4 Turbo (9-shot) & 17 & 24 & 20 & 34 & 54 & 42 & 30 & 4 & 7 & 23 \\
		\midrule
		Llama3 8B with RAG & 23 & 28 & 26 & 43 & 51 & 47 & 53 & 37 & 44 & 39 \\
		Llama3.3 70B with RAG & 22 & 51 & 30 & 34 & 22 & 27 & 62 & 32 & 42 & 33\\
		GPT-3.5 Turbo with RAG & 19 & 8 & 11 & 42 & 71 & 53 & 60 & 40 & 48 & 37 \\
		GPT-4 Turbo with RAG & 15 & 34 & 21 & 31 & 33 & 32 & 55 & 11 & 19 & 24 \\
		\midrule
		\citet{Cruciani:ScalableTestSuiteReduction:ICSE:2019} & 32 & 32 & 32 & 27 & 27 & 27 & 44 & 44 & 44 & 34 \\ 
		\citet{Gazzola:FieldReadyTesting:ICST:2022} & 50 & \textbf{100} & 67 & 0 & 0 & 0 & 18 & 33 & 23 & 30 \\\midrule
		\approach & \textbf{66} & 94 & \textbf{78} & 47 & 24 & 31 & 48 & \textbf{57} & \textbf{53} & \textbf{54} \\
		\approach with RAG & 47 & 41 & 44 & \textbf{51} & 61 & \textbf{56} & \textbf{52} & 46 & 49 &  50 \\
		\bottomrule
	\end{tabular}
    }
	\label{tab:effectiveness_metrics}
\end{table}

\section{Contributions}

This paper contributes to software testing by proposing
\begin{enumerate*}[label=(\roman*)]
    \item a novel viewpoint for long-lasting test suite augmentation using untested execution scenarios,
    \item \approach, an effective approach to classify unit execution scenarios, and generate failure-revealing test cases using LLMs,
    \item a dataset of \allScenarios unit execution scenarios that we created from popular Java projects,
    \item a solid experimental evaluation of \approach with widely-used LLMs under different settings and comparison with two state-of-the-art approaches.
\end{enumerate*}

\begin{acks}
  This work is supported by the \grantsponsor{SNF}{Swiss SNF project A-Test Autonomic Software Testing}{} ({SNF 200021\_215487}).

\end{acks}

%%
%% The next two lines define the bibliography style to be used, and
%% the bibliography file.
\balance
\bibliographystyle{ACM-Reference-Format}
\bibliography{fse25src.bib}

\end{document}